\documentclass[floatfix,aps,prb,twocolumn,showpacs]{revtex4}
\usepackage[T1]{fontenc}

\usepackage{graphics}
\usepackage{epsfig}
\usepackage{psfig}
\newcommand{\beq}{\begin{equation}}
\newcommand{\eeq}{\end{equation}}
\newcommand{\bea}{\begin{eqnarray}}
\newcommand{\eea}{\end{eqnarray}}
\newcommand{\nn}{\nonumber}

\newcommand{\eps}{\epsilon}
\newcommand{\veps}{\varepsilon}
\newcommand{\al}{\alpha}

\newcommand{\D}{\Delta}

\newcommand{\be}{\beta}

\newcommand{\ra}{\rangle}
\newcommand{\la}{\langle}

\newcommand{\ga}{\gamma}
\newcommand{\om}{\omega}

\newcommand{\app}{\approx}

\begin{document}
\title{Transport Spectroscopy of a Kondo quantum dot coupled to a finite size grain.}

\author{Pascal Simon$^1$, Julien Salomez$^1$ and Denis Feinberg$^2$}

\affiliation{
$^1$Laboratoire de Physique et Mod\'elisation des Milieux
     Condens\'es, CNRS et Universit\'e Joseph Fourier, 38042 Grenoble, France \\
$^2$
 Laboratoire d'Etudes des Propri\'et\'es Electroniques
des Solides, CNRS, associated with Universit\'e Joseph Fourier, BP 166, 38042 Grenoble, France}
\date{\today}

\begin{abstract}
We analyse a simple setup in which a quantum dot is strongly connected to a metallic grain or finite size wire and weakly connected to two normal leads. The Kondo screening 
cloud essentially develops in the strongly coupled grain
whereas the two weakly connected reservoirs can be used as transport probes.
Since the transport channels and the screening channels are almost decoupled,
such a setup allows an easier access to the measure of finite-size Kondo 
effects. 
\end{abstract} 

\maketitle
\section{Introduction}
\label{intro}

One of the most remarkable achievements of recent progress in nanoelectronics 
has been the observation of the Kondo effect in a single semi-conductor 
quantum dot. \cite{dot,Cronenwett,Wiel}   When the number of electrons in 
the dot is odd, it can behave as an $S=1/2$ magnetic impurity interacting via 
magnetic exchange with the conduction electrons. One of the main signatures
of the Kondo effect is a zero-bias anomaly and the conductance reaching the unitary
limit $2e^2/h$ at low enough temperature $T<T_K^0$. $T_K^0$ stands for the 
Kondo temperature and is the main energy scale of the problem. 
At low temperature, the impurity spin is screened and 
forms a singlet with a conduction electron belonging to a very extended many-body
wave-function known as the Kondo screening cloud. 
The 
size of this screening cloud may be evaluated as $\xi_K^0\approx \hbar v_F/T_K^0$ where 
$v_F$ is the Fermi velocity. In a quantum dot, the typical Kondo temperature is of order
$1~K$ which leads to $\xi_K^0\approx 1$ micron in semiconducting heterostructures.
Finite size effects (FSE) related to the actual extent of this length scale have been 
predicted
recently in  different geometries: an impurity embedded in a finite size box,\cite{thimm}
a quantum dot embedded in a ring threaded by a magnetic flux,
\cite{affleck01,Hu,sorensen04,simon05} 
and a quantum dot embedded between two open finite size wires (OFSW) (by open we mean 
connected to at least one external infinite lead).\cite{simon02,balseiro}
In the ring geometry, it was shown that the persistent 
current induced by a magnetic flux is particularly sensitive to screening 
cloud effects and is drastically reduced when the circumference 
of the ring becomes smaller than $\xi_K^0$.\cite{affleck01} In the wire geometry, 
a signature of the finite size extension  of the Kondo cloud was found in the temperature dependence of the 
 conductance through the whole system.\cite{simon02,balseiro} To be more precise, in a one-dimensional 
 geometry where the finite size $l$ is associated to a level spacing $\D$, the Kondo cloud fully develops 
 if $\xi_K^0 \ll l$, a condition equivalent to $T_K^0 \gg \D$. On the contrary, FSE effects appear if 
 $\xi_K^0 > l$ or $T_K^0 < \D$.

Nevertheless, in such a two-terminal geometry, the screening of the artificial spin impurity is done in the OFSWs which are also used to probe
transport properties through the whole system. This brings at least two main drawbacks: 
first, the analysis of FSE relies on the independent 
control of the two wire gate voltages and also a  rather symmetric geometry. This 
is difficult to achieve experimentally. 
In order to remedy these drawbacks, we propose and study here a simpler setup in which 
the screening of the impurity occurs mainly in one larger quantum dot or metallic grain\cite{note0} or OFSW 
and the transport is analyzed by help of one or two weakly coupled leads. 
In practice, a lead weakly coupled to the dot by a tunnel junction allows a 
spectroscopic analysis of the dot local density of states (LDOS) in a way very similar to a STM tip.
The rather general geometry we study is
depicted in Fig. \ref{Fig:device}. We note that this geometry has also been proposed by Oreg and
Goldhaber-Gordon \cite{oreg} to look for signatures of the two-channel Kondo fixed point or by Craig et al. \cite{craig} to analyze two quantum dots coupled to a common larger quantum dots and interacting {\it via} 
the RKKY interaction.
In the former case, the key ingredient is the Coulomb interaction of the grain whereas in the latter, the grain is largely open and used simply as a metallic reservoir mediating both the Kondo and RKKY interactions.
Here we are interested in the case where finite-size effects in the larger quantum dot or metallic
grain do matter whereas in the aforementioned experiments the level spacing was among the smallest scales.

The plan of the paper is the following: in section 2, we present the model Hamiltonian
and derive how the FSE renormalizes
the Kondo temperature in our geometry.
In section 3, we show how FSE affect the transport properties of the quantum dot
and perform a detailed spectroscopic analysis. The effect of a finite Coulomb energy in the grain is also discussed. Finally section 4 summarizes our results.

\section{Presentation of the model}
\label{sec:1}
\subsection{Model Hamiltonian}

\begin{figure}
\epsfig{figure=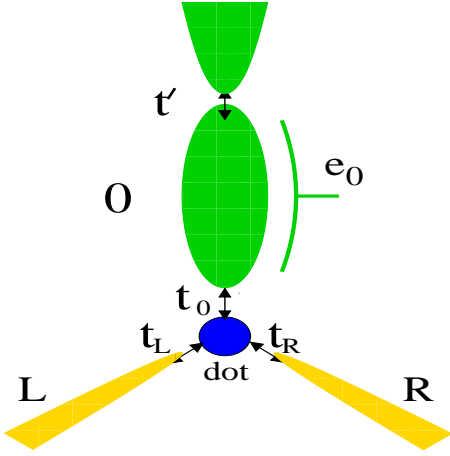, width=6.cm,height=6.cm}
\caption{Schematic representation of the device under analysis. When the grain Coulomb blockade energy is not neglected, 
we assume that the grain potential can be controlled by a voltage gate $e_0$.}
\label{Fig:device}
\end{figure}
The geometry we analyze is depicted in Fig. \ref{Fig:device}. In this section we
assume that the large dot is connected to a third lead. We neglect the Coulomb interaction
in the grain. As we will see in the next section, the Coulomb interaction does not affect
the main results we discuss here.
In order to model the finite-size grain connected to a normal reservoir, we choose for convenience
a finite-size wire characterized by its length  $l$ or equivalently by its level spacing $\Delta\sim \hbar v_F/l$. 
In fact, the precise shape of the finite-size grain is not important for our purpose as soon as it is 
characterized by  a mean level spacing $\Delta$ separating peaks in the electronic density of states.
We assume that the small quantum dot is weakly coupled to one or two adjacent leads ($L$ and $R$). 
On the Hamiltonian level, we use the following tight-binding description, and for simplicity 
model the leads as one-dimensional wires (this is by no means restrictive): $H=H_L+H_R+H_0+H_{dot}+H_{tun}$ with
\bea\label{hamil}
H_{L}&=&-t\sum_{j=1,s}^{\infty} (c^\dagger_{j,s,L}c_{j+1,s,L}+h.c.)
-\mu_{L}  n_{j,s,L}  \\
H_0&=& -t\sum\limits_{j=1,s}^{\infty} (c^\dagger_{j,s,0}c_{j+1,s,0}+h.c.)-\mu_0 n_{j,s,0}\\
&&+(t-t')\sum_s (c^\dagger_{l,s,0}c_{l+1,s,0}+h.c.)\nn\\
H_{dot}&=&\sum_s \epsilon_{d} n_{d,s} +Un_{d\uparrow}n_{d\downarrow}\\
H_{tun}&=&\sum_s\sum_{\al=L,R,0}( t_\al c^\dagger_{ds} c_{1,s,\al}+h.c.).
\eea
$H_R$ is obtained from $H_L$ by changing $L\to R$.
Here $c_{j,s,\al}$ destroys an electron of spin $s$ at site $j$ in lead 
$\al=0,L,R$; $c_{d,s}$ destroys an electron with spin $s$ in the dot, $ n_{j,s,\al}= c^\dagger_{j,s,\al}c_{j,s,\al}$ and $n_{ds}= c^\dagger_{ds}c_{ds}$. The quantum dot is described by an Anderson impurity model, $\eps_{d},U$ are respectively the energy level and the Coulomb repulsion energy in the dot.
The tunneling amplitudes between the dot and the left lead, right lead and grain are respectively 
denoted as $t_L,t_R,t_0$ (see Fig. 1). The tunneling amplitude amplitude between the grain and the third lead is denoted
as $t'$ (see Fig. 1). Finally $t$ denotes the tight binding amplitude for conduction electrons 
implying that the electronic bandwidth $\Lambda_0=4t$. 
Since we want to use the left and right leads just as transport probes, we
assume in the rest of the paper that $t_L,t_R \ll t_0$.

We are particularly interested in the Kondo regime where $\la n_d\ra\sim 1$.
In this regime, we can map $H_{tun}+H_{dot}$ to a Kondo Hamiltonian by help of a Schrieffer-Wolff transformation:
\beq\label{hkondo}
H_K=H_{tun}+H_{dot}=\sum_{\alpha,\beta=L,R,0}
J_{\alpha\beta}c_{1,s,\alpha}^{\dagger}\frac{\vec{\sigma}_{ss'}}{2}\cdot\vec{S}c_{1,s',\beta},
\end{equation}
where $J_{\al\beta}=2t_\al t_\be(1/|\veps_d|+1/(\veps_d+U))$.
It is clear that $J_{00}\gg J_{0L},J_{0R}\gg J_{LL},J_{RR},J_{LR}$.
In Eq. (\ref{hkondo}), we have neglected direct potential scattering terms
which do  not renormalize and can be omitted in the low energy limit.

\subsection{Kondo temperatures}
The Kondo temperature is a crossover scale separating the high temperature perturbative regime from the low temperature 
one where the impurity is screened.
There are many ways to define such scale. We choose the ``perturbative scale'' which is defined as the scale at which the 
second order corrections to the Kondo couplings become of the same order 
of the bare Kondo coupling. Note that all various definitions of Kondo scales differ by a constant multiplicative facor (see for example Ref. [\onlinecite{simon05}] for a comparison of the perturbative Kondo scale with the one coming from the Slave Boson Mean Field Theory).

The renormalization group equations relate the Kondo couplings defined at scales $\Lambda_0$ and $\Lambda$. They simply 
read:
\bea \label{RG0}
J_{\al\be}(\Lambda)&\approx& J_{\al\be}(\Lambda_0)\\
&+&{1\over 2}\sum_\ga J_{\al\ga}(\Lambda_0)J_{\ga\be}(\Lambda_0)
\left[\int\limits_\Lambda^{\Lambda_0}+\int\limits_{-\Lambda_0}^{-\Lambda}\right] {\rho_\ga(\om)\over |\om|}d\om\nn
\eea
where $\rho_\gamma$ is the LDOS in lead $\ga$ seen by the quantum dot. 
When the density of states $\rho_\ga$ are uniform, the RG equations can be rewritten
 by introducing  $\hat \lambda$, the matrix of the dimensionless Kondo couplings: 
$\lambda_{\al\be}=\sqrt{\rho_\al\rho_\beta}J_{\al\be}$. \cite{simon02,note1}
The Kondo temperature $T_K$ may be therefore defined as
\beq
{1\over 2} \left[\int\limits_{T_K}^{\Lambda_0}+\int\limits_{-\Lambda_0}^{-T_K}\right] {Tr(\hat\lambda(\om))\over |\om|}d\om=1.
\eeq

Since $J_{00}\gg J_{LL},J_{RR}$, the Kondo temperature essentially depends on the LDOS in the lead $0$ and the Kondo 
temperature definition can be well approximated by
\beq\label{deftk}
{ J_{00}\over 2}\left[\int\limits_{T_K}^{\Lambda_0}+\int\limits_{-\Lambda_0}^{-T_K}\right] {\rho_0(\om)\over |\om|}d\om=1
\eeq
When the lead $0$ becomes infinite (i.e. when $t'=t$), $\rho_0(\om)=\rho_0=const$ and we recover $T_K=T_K^0$
the usual Kondo temperature where a constant density of states is assumed.
It is worth noting that including the Coulomb interaction in the grain does not affect much the Kondo temperature. 
The grain
Coulomb energy $E_G$ slightly renormalizes $J_{00}$ in the Schrieffer-Wolff transformation\cite{oreg}since 
$E_{G}\ll U,|\eps_d|$.

The LDOS $\rho_0$ can be easily computed and corresponds in the limit $t'^2\ll t^2$
to a sum of peaks at positions $\om_n=-2t\cos k_{0,n}-\mu_0$ of width
$\gamma_{n} \approx {2(t')^2\sin^3(k_{0,n})\over t l}$
where $
k_{0,n}\app \pi n/(l+1)+O(t'^2/t^2 l)$ (Ref. \onlinecite{simon02}). The LDOS $\rho_0$
is very well approximated by a sum of Lorentzian functions\cite{simon02} in the limit $t'\ll t$ :
\beq\label{lorentzienne}
\pi \rho_0(\om)\approx \frac{2}{l+1}\sum_n \sin^2(k_{0,n}) \frac{\gamma_n}{(\om-\om_n)^2+\gamma_n^2}.
\eeq
This approximation is quite convenient in order to estimate the Kondo temperature $T_K$ through (\ref{deftk}). 
Note that the position of the resonance peaks may be controlled by the chemical potential $\mu_0$.
Another possibility is to fix $\mu_0$ and to add a small transverse magnetic field which modifies the orbital
part of the electronic 
wave function in the grain and therefore shift the resonance peak positions. 
When the level spacing $\Delta_n \sim 2\pi t \sin k_{0,n}/l$ is much smaller
than the Kondo temperature $T_K^0$, no finite-size effects are expected. Indeed, the integral in (\ref{deftk})
 averages out over many peaks and the genuine Kondo temperature is $T_K\sim T_K^0$.
On the other hand, when $T_K^0 \sim\Delta_n $, the Kondo temperature begins to depend on the fine structure 
of the LDOS $\rho_0$ and a careful calculation of the integral in
(\ref{deftk}) is required. Two cases may be distinguished: either $\rho_0$ is tuned such that 
a resonance  $\om_n$ sits at the Fermi energy $E_F=0$ (labeled by the index $R$) or in a non resonant situation  
(labeled by the index $NR$). 

In the former case, we can estimate
\bea\label{def:tkr}
T_K^R&=&\frac{\gamma_n \Delta_n}{\sqrt{(\Delta_n^2+\gamma_n^2)\exp\left(\frac{2}
 {J_{00}(\Delta_n )\rho_0^R(0)}\right)-\Delta_n^2}}\nn \\
&\approx&
\gamma_n\exp\left(-\frac{1}
 {J_{00}(\Delta_n)\rho_0^R(0)}\right)\approx \ga_n\left(\frac{T_K^0}{\D_n}\right)^{\frac{\pi\ga_n}{\D_n}},
\eea
where we approximate the on-resonance LDOS at $E_F=0$ by $\pi\rho_0^R(0)\approx 2\sin^2 k_{0,n}/l\gamma_n$.
In the latter case, we obtain,
\bea\label{def:tknr}
T_K^{NR}&=&\Delta_n \exp\left(-\frac{\pi l\Delta_n^2}{16 J_{00}(\Delta_n)\sin^2(k_{0,n})\gamma_n} \right)\nn\\
&\approx& \Delta_n \exp\left(-\frac{1}{J_{00}(\Delta_n)\rho_0^{NR}(0)}\right)
\approx  \Delta_n \left(\frac{T_K^0}{\D_n}\right)^{\frac{\pi\D_n}{8\ga_n}}, 
\eea
where $\pi\rho_0^{NR}(0)\approx 16\gamma_n\sin^2 k_{0,n}/l\Delta_n^2$.
These two scales are very different when $t'^2\ll t^2$. By controlling $\rho_0$, we can control the 
Kondo temperature (only when $T_K\leq \D$).
The main feature of such geometry is that the screening of the artificial spin impurity is essentially 
performed in the open finite-size wire corresponding to lead $0$. Now we want  to study what are the consequences of FSE  on transport 
when one or two leads are weakly coupled to the dot. This is the purpose of the next section.

\section{Transport spectroscopy of a quantum dot coupled to an open grain}\label{sec:spectro}

In this section, we consider a  standard three-terminal geometry as depicted in Fig. \ref{Fig:device} and
analyze the conductance matrix of the system. 
There are several approaches we may combine to obtain such quantity
for the whole temperature range. Nonetheless, before going into these details, let us analyze the dot density
of states in presence of FSE in the grain.

\subsection{Density of states}
We  have used the Slave Boson Mean Field
Theory (SBMFT) \cite{hewson}  in order to calculate the dot density of states.
This approximation describes qualitatively well the behavior of 
the Kondo impurity
at low temperature $T\leq T_K$ when the impurity is screened. 
This method has been proved to be efficient to capture finite size effects in Ref. \onlinecite{simon02}.
The main advantage of the SBMFT relies on its ability to qualitatively reproduce  the energy scales of the problem  
(here the Kondo temperature).

\begin{figure}
\psfig{figure=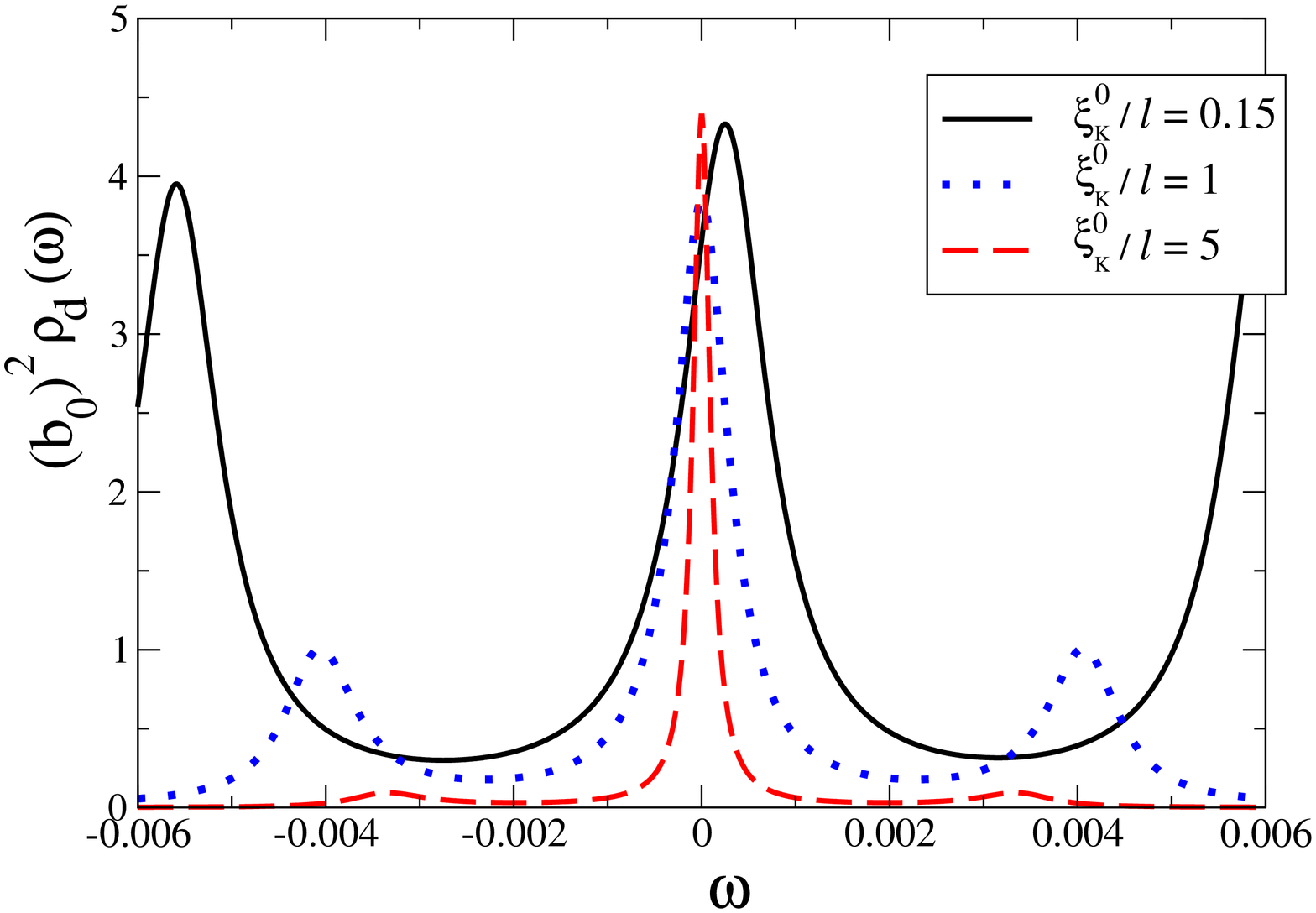, width=8.cm,height=8.5cm,angle=-90}
\psfig{figure=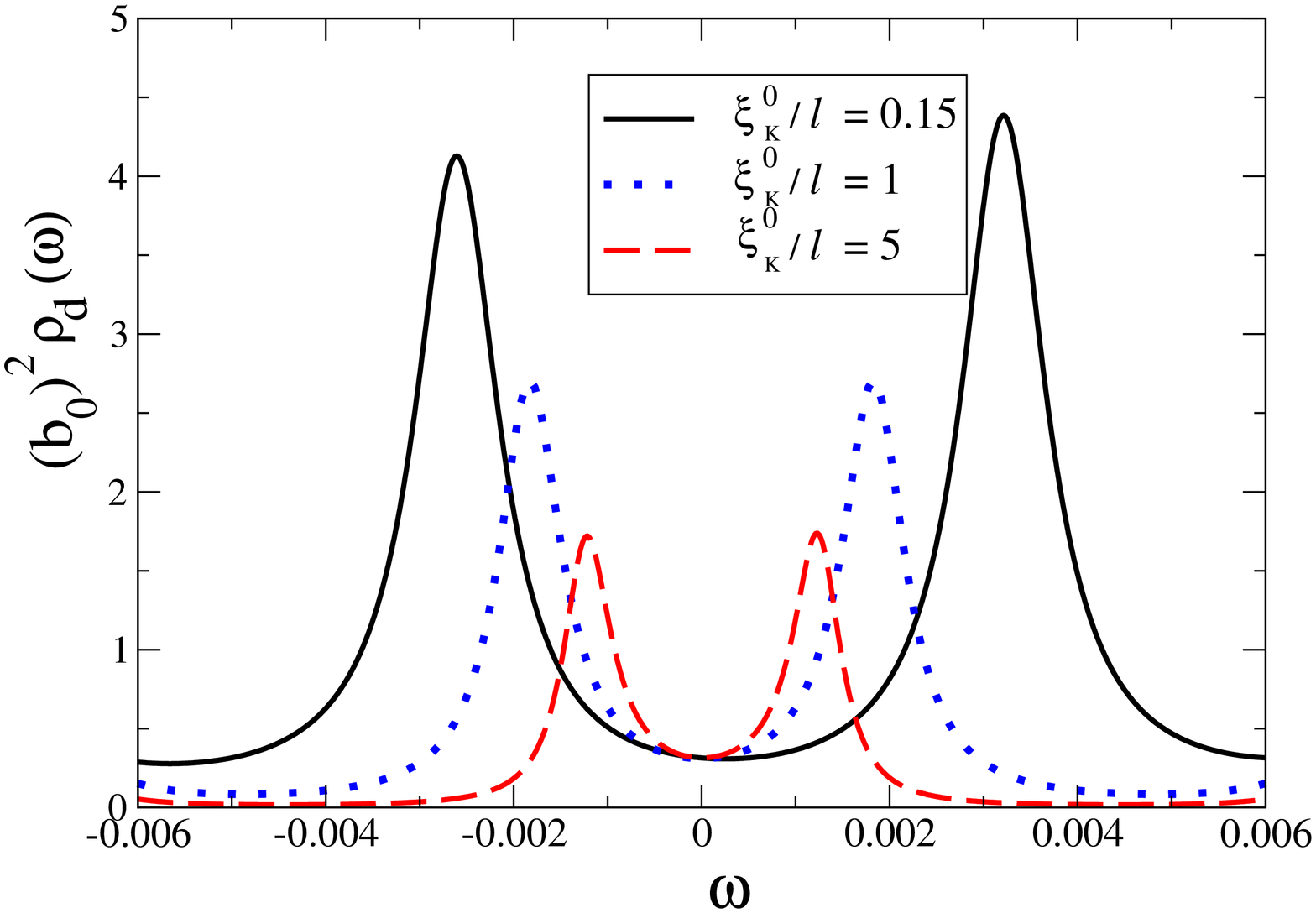, width=8.cm,height=8.5cm,angle=-90}
\caption{Dot density of states $\rho_d(\omega)$ for  both the non resonant case (upper panel)
and the resonant case (lower panel). We took $\Delta\sim 0.006$ and  plot $\rho_d$ for $\xi_K^0/l=\Delta/T_K^0\sim 0.15 $, $\xi_K^0/l=\Delta/T_K^0\sim 1$ and $\xi_K^0/l=\Delta/T_K^0\sim 5$. 
Note that $\rho_d$ has been scaled by $b_0^2$, the slave boson parameter for an easy comparison between both cases. }
\label{Fig:rhod}
\end{figure}

An interesting quantity to look at is the dot density of states $\rho_d(\omega)$.
The density of states can be read from the differential conductance as follows:
 The current in the left electrode $I_L$ reads:\cite{meir}
\beq
I_L=\frac{4e}{h}\Gamma_L\int\limits_{-\infty}^\infty\left[2\pi f(\om-\mu_L)\rho_d(\om)-G_d^<(\om)\right]d\om,
\eeq
where $\rho_d$ is the dot LDOS, $G_d^<(\om)=\int e^{i\om t}\la d^\dag d(t)\ra$ the lesser Green function for the dot and $\Gamma_{L/R}=\pi t_{L/R}^2\rho_{L/R}$.
The standard procedure is to get rid of the lesser dot Green function using the current conservation 
(here $I_L+I_R+I_0=0$). Nevertheless, such procedure is useful only when the leads density of states can be
assumed constant on the typical scale we are interested in. This is not the case here 
because of the variations
of $\Gamma_0(\om)=\pi t_0^2\rho_0(\om)$. Nevertheless, one can make progress 
by assuming $\Gamma_{L/R}\ll \Gamma_0(\om)$ such that for low bias the dot Green functions weakly depend on
the chemical potential in the left and right leads.
Therefore
\beq
e\frac{dI}{d\mu_L}\approx\frac{2e^2}{h}4\Gamma_L\int\limits_{-\infty}^\infty\left(\frac{-df(\om)}{d\om}\right)
\pi\rho_d(\om+\mu_L) d\om,
\eeq
which allows an experimental access to $\rho_d(\mu_L)$.
Note that a similar approximation is used for STM theory with magnetic adatoms \cite{schiller00}.
We have plotted $\rho_d(\omega)$ in Fig. \ref{Fig:rhod} for both the non-resonant case
and the on-resonance case for three different values of
$\xi_K^0/l$. We took the following parameters in units of $t=1$: $t_0=0.5$, 
$t_L=t_R=0.1$ (therefore $t_L^2<<t_0^2$), $t'=0.5$, and $l\sim 1000a$ ($a$ the lattice constant) or
equivalently $\Delta\sim 0.006$. 
The Kondo energy scale $\xi_K^0$ can be varied using the dot energy level $\eps_d$ which is controlled 
by the dot gate voltage.  When $T_K^0 \gg \D$, no finite-size effect is to be expected. 
In this case, $\rho_d(\omega)$
mimics the density of state in the lead $0$ but is shifted  such that an off-resonance peak
in the lead  $0$ corresponds to a dot resonance peak.
The various peaks appearing in $\rho_d$ are included in an envelope of width $O(T_K^0)\gg \Delta$
(which is a broader range than the figure \ref{Fig:rhod} actually covers for $\xi_K^0/l=\Delta/T_K^0\sim 0.15$). 
This can be simply understood from a non-interacting picture valid at $T=0$. The non-interacting dot Green function 
reads 
\beq
G_{dd}(\omega)\approx\frac{1}{\omega -\eps_d -\delta \eps(\omega)+i\Gamma_0(\omega)},\eeq
where $\delta\eps$ is the real part of the dot self-energy and $\Gamma_0$ its imaginary part.
The minima's of $\Gamma_0$ thus correspond to the maxima of $-Im(G_{dd})$. 
We also note that the resonance peak
is slightly shifted from $\om=0$ in this limit. Therefore the 2-terminal conductance does not reach its unitary limit (i.e. its maximum non interacting value). 
This is due to the fact that we took $\eps_d=-0.68$ and we are not deep in the Kondo regime.
Particle-hole symmetry is not completely restored in the low energy limit.
 
On the other hand, when $\xi_K^0\gg l$ ($T_K^0\ll \D$), $\rho_d$ changes drastically. 
The fine structure in the grain density of states no longer shows up in $\rho_d$. 
In the off-resonance case, only the narrow Kondo 
peak of width $T_K^{NR}\ll T_K^0$ mainly subsists for $\xi_K^0/l=5$ (upper panel of Fig. \ref{Fig:rhod}).
 We can also show that the position of the small peaks at $\omega\sim \pm\Delta/2$ for $\xi_K^0\gg l$ 
are related to the resonance peaks in lead $0$. We also note that the narrow peak is this time almost at $\om=E_F=0$
i.e. particle-hole symmetry is restored. In order to reach large value of $\xi_K^0$, we took small values
of $\eps_d$ such that we are deep in the Kondo regime where $n_D\sim 1$.

In the resonant case, the Kondo peak is split for $\xi_K^0\gg l$. By approximating 
\beq\label{gamma0}
\Gamma_0(\omega)\approx t_0^2 \sin^2(k_{0,n})\frac{\gamma_n/l}{\omega^2+\gamma_n^2},\eeq
 at a resonance $n$, 
one can well understand
the structure of $\rho_d$ with the SBMFT. When $T_K^R\gg \ga_n$, 
one can  show that the peak splitting is of order 
$\sim 2\sqrt{\gamma_n T_K^R}$ and the peaks width is of order $\gamma_n$. 
We have shown that the dot density of states $\rho_d(\om)$ has interesting features
at $T=0$ where FSE can be clearly identified. Let us summarize them:  First, when $T_K^0\gg \Delta$, $\rho_d$ has several peaks embedded in an enveloppe of width $\sim T_K^0$ associated with the grain one-particle levels and one or two Kondo peaks in the opposite limit $T_K^0\ll \Delta$.
Second, the position of the peaks and their widths depend on the grain being on- or off- resonance.

\subsection{Finite temperature conductances}
In this section, we analyze the conductance matrix using various methods. 
We start with the low temperature limit.

\subsubsection{Low temperature regime}

\begin{figure}
\psfig{figure=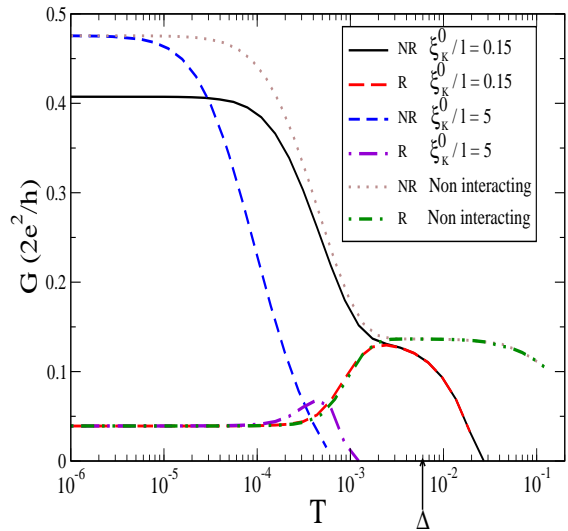, width=8.5cm,height=8.5cm,angle=-90}
\caption{Conductance between lead $L$ and lead $0$ (in units of $2e^2/h$) as a function of temperature when the density of states in the grain  is tuned on resonance (curves denoted as R) and out of resonance (curves denoted as NR). 
Three cases were considered: $\xi_K^0/l\sim 0.15 $ (lower 
plain style and upper long dashed curves), $\xi_K^0/l\sim 5$ (lower dashed and upper dot-dashed curves). We compare these curve to the non interacting case corresponding to $\eps_d=U=0$ (dotted lines). The level spacing $\D$ is indicated by an arrow on the x axis.}\label{Fig:conducl0}
\end{figure}

At $T=0$, it is straightforward to show, using for example the scattering formalism,\cite{ng} 
that the conductance matrix $G_{\al,\be}^U$ is simply given
by

\beq\label{giju}
G_{\al,\be}^U=\frac{2e^2}{h} {4\Gamma_\al\Gamma_\be\over (\Gamma_L+\Gamma_0+\Gamma_R)^2}
\eeq
where $\Gamma_\al=\pi t_\al^2\rho_\al(0)$, and $\al=l,0,r$.
Since the SBMFT aims at replacing the initial Anderson Hamiltonian by a non-interacting one,
one may easily access the conductance by directly applying the Landauer formula
or equivalently by using
\beq\label{mw1}
G_{\al\be}=\frac{2e^2}{h}\int d\omega \left(\frac{-\partial f}{\partial \omega}\right) 
 {4\Gamma_\al(\omega)\Gamma_\be(\omega)\over (\sum_\al \Gamma_\al(\omega))}
Im(-G_{dd}^r)(\omega).
\eeq
We have plotted in Fig. \ref{Fig:conducl0}, the conductance between the left lead $L$ and the lead $0$
for both the on- and off-resonance cases. Note that in absence of Coulomb energy the density of states in
the grain can be controlled by the chemical potential $\mu_0$.

The three upper curves correspond 
to the non-resonant cases whereas the three lower curves correspond 
to the resonant cases. In the standard case corresponding to $\xi_K^0\ll l$ ($T_K^0\gg \Delta$), the conductance 
qualitatively follows
the non interacting limit except at high temperature (here $T\gg T_K^0\sim 0.012$) 
where the conductance falls down  in the Kondo regime. 
The conductances corresponding to $\xi_K^0\gg l$ 
are characterized by an abrupt increase of the 
conductance at their respective Kondo temperatures $T_K^R$ and $T_K^{NR}$ which are lower 
than $T_K^0$.

The main signature of finite size effects occurs at an intermediate temperature $T_K^0> T\sim 0.001>T_K^R,T_K^{NR}$. In this temperature range, the conductance is already  a fraction of its unitary limit value when 
$\xi_K^0/l\sim 0.15 \ll 1$ whereas it
is still very low when $\xi_K^0/l\sim 5 >1$. In this temperature range, 
significant deviations from 
the non-interacting limit are obtained for $\xi_K^0/l\sim 5$  contrary 
to the case where $\xi_K^0/l\sim 0.15$ .
We also notice that the conductance for the on-resonance case is non monotonous. This is analyzed and confirmed analytically further using Nozières' 
Fermi liquid approach.
As anticipated before from the spectroscopic analysis, the unitary limit is not fully  reached for the NR case at $\xi_K^0<l$.
It corresponds to the fact that particle-hole symmetry is not fully restored at low energy 
in the SBMFT approximation. 
Because of the large value of $T_K^0$ we took (in order to satisfy $\xi_K^0<l$) the dot occupation $n_d$
starts slightly deviating from $1$ which explains why the unitary limit is not fully restored at low $T$. 

Similar results can be obviously obtained by analyzing the 
conductance $G_{LR}$ despite the fact that 
the amplitude of $G_{LR}$ will be in general smaller than $G_{L0}$ by a factor $\sim \sqrt{\Gamma_L\Gamma_R}/\Gamma_0$. In fact, the conductance $G_{LR}$ can be made significantly larger by adjusting the chemical potential $\mu_0$ such that $\langle I_0 \rangle=0$. 
The grain or OFSW density of states $\rho_0$ can then be controlled by a small orbital magnetic field or by gating the grain.
In the linear response regime, 
this can always be realized provided
$\mu_0=(G_{L0}\mu_L+G_{R0}\mu_R)/(G_{L0}+G_{R0})$. In this sense, the geometry becomes a two-terminal one.
More generally, one  can  also show using a general non-equilibrium approach following Ref. [\onlinecite{meir}] 
that there is a unique value of $\mu_0$ such that $I_0=0$.
Note that for a symmetric device (between left and right) and $\mu_L=-\mu_R=eV/2$, 
symmetry considerations imply $I_0=0$ for $\mu_0=0$. 
For voltages smaller than the smallest energy scale of the problem ($T_K^0$ or $\gamma_n$), we can safely use
the simplified Meir-Wingreen formula.\cite{meir} This ``two-terminal like '' conductance then reads:
\beq\label{mw}
G_{LR}^{(2)}=-\frac{2e^2}{h}\int d\om \left(\frac{-\partial f}{\partial \om}\right) \Gamma_{LR}(\om)
Im(G_{dd}^r)(\om),
\eeq
where 
$\Gamma_{LR}={4\Gamma_L\Gamma_R\over (\Gamma_L+\Gamma_R)}$. We have added the index $(2)$ to distinguish it from 
the $3$-terminal conductance.
Contrary to Eq. (\ref{mw1}), this equation is exact
and does not rely on a mean-field approximation.
 Under the approximation $t_L,t_R\ll t_0$, one may relate the exact dot Green's function to the on shell $T$-matrix
associated with the scattering problem of an electron in lead $0$ by the dot: $T(\veps)\approx t_0^2 G_{dd}^r$ 
such that:
\beq\label{tmatrix}
G^{(2)}_{LR}\approx\frac{2e^2}{h}\int d\om \left(\frac{-\partial f}{\partial \om}\right) 
{\Gamma_{LR}(\om)\over \Gamma_0(\om)} \left(-\pi\rho_0(\om) Im(T(\om))\right).
\eeq
We see that by adjusting $\mu_0$  such that $I_0=0$, the amplitude of $G_{LR}^{(2)}$  can be made of the order of $G_{L0}$.
We have checked that $G_{LR}^{(2)}(T)$ has the same features as $G_{L0}(T)$ as it should be.

\subsubsection{Nozières' Fermi liquid approach}
Equation (\ref{tmatrix}) may be good starting point to analyze the finite-temperature Fermi liquid correction 
to the unitary limit.
In the Fermi liquid approach the on-shell $T$-matrix is expanded at low $T$ as a function
of $\om$ and $T$. Here new energy scales are introduced and this expansion must be done with care.
Let us start with the off-resonance case where at low energy $\rho_0(\om)\sim const$.
In this case, one can safely use the usual $T$- matrix expansion\cite{nozieres}
\beq\label{t_matrix}
-\pi \rho_0(\om) Im(T(\om))\approx 1-\frac{3\om^2+\pi^2T^2}{2(T_K^{NR})^2}.\eeq

We then recover the standard Fermi liquid corrections  
\beq
G^{(2)}_{LR}(T)\approx G^{(2)}_{U}\left(1-\pi^2\left(T\over T_K^{NR}\right)^2\right)~,
\eeq 
where $G^{(2)}_{U}=4\Gamma_L\Gamma_R/(\Gamma_0(\Gamma_L\Gamma_R))$ .
Therefore the off-resonance conductance always decreases at low temperature as depicted in Fig. \ref{Fig:conducl0}.
However the situation is more subtle when $\rho_0$ is  on a resonance. 

In order to perform an expansion of the on-shell T-matrix analogous to Eq. (\ref{t_matrix}),
one should first pay attention to the new energy scale $\gamma_n$. 
When $\gamma_n\gg T$, one may reproduce the same
expansion as in [\onlinecite{nozieres}]. Eq. (\ref{t_matrix}) remains qualitatively valid provided
we replace $T_K^{NR}$ by $T_K^R$.
Reporting together both Eqs (\ref{t_matrix}) and (\ref{gamma0})
in Eq. (\ref{tmatrix})
and performing a Sommerfeld expansion, we find:
\beq\label{glowt}
G^{(2)}_{LR}(T)\approx G^{(2)}_{U}
\left(1+\pi^2 T^2\left(\frac{ 2 }{3(\gamma_n)^2}-\frac{1}{(T_K^R)^2}\right)\right),
\eeq
We find two opposite contributions to the conductance at low temperature:
the usual negative one coming from the $T$-matrix expansion and a second positive one depending on the
resonance width. The latter has nothing to do with interactions and is simply an effect of quantum 
interference 
analogous to the Fano resonance recently seen in quantum dot experiments.\cite{fanodot} For a narrow resonance satisfying $\gamma_n \ll T_K$, the conductance
increases at low $T\ll \gamma_n$. On the other hand, for a broad resonance, we recover the usual situation.
Note that this value suggests an intermediate value of $\gamma_n$ where the $T^2$ corrections cancel.
For the 3-terminal conductance matrix, we can perform a similar analysis starting from Eq. (\ref{mw1}) which can be regarded as a good approximation
of the conductance in Nozières' Fermi liquid regime and obtain almost similar conclusions.

\subsubsection{High temperature}
  At temperature higher than the Kondo temperature $T\gg T_K^0$, 
perturbation theory in the Kondo couplings can {\it a priori} be used. 
We therefore start  directly from  the Kondo Hamiltonian written in Eq. (\ref{hkondo}) 
In order to compute the conductance between the left lead and lead 0, 
we use the renormalized perturbation theory. 
At lowest order, the conductance reads:
\beq
G_{L0}=\frac{2e^2}{h} \frac{3\pi^2}{4}J_{L0}^2\int d\om \left(\frac{-\partial f}{\partial \om}\right)\rho_L\rho_0(\om) 
\eeq
When $T_K^0\gg \Delta$, one may replace $\rho_0$ by its average value and $J_{L0}$ by its renormalized coupling and the high temperature conductance takes its standard scaling form\cite{glazman05}
\bea \label{conduc1}
G_{L0}&=&G^U_{L0} \frac{3\pi^2/16}{\ln^2(T/T_K^0)}~~,T\gg T_K^0\gg\Delta\\
&=& G^U_{L0} f(\frac{T}{T_K^0})\nn,\eea
where $f(x)$ is a universal scaling function such that $f(x\gg 1)\approx \frac{3\pi^2}{16\ln^2x}$
and $G_{L0}^U $ has been defined in Eq. (\ref{giju}).

Let us consider the more interesting situation where $T_K^0<\Delta$.
Suppose the  bandwidth is reduced from $\pm \Lambda_0$ (where $\lambda_0=4t$ here)
to $\pm \Lambda$, the renormalization of the Kondo coupling is well approximated by: 
\beq\label{rg}
J_{L0}(\Lambda)=J_{L0}(\Lambda_0)+\frac{1}{2}J_{L0}J_{00}\left[\int\limits_{-\Lambda_0}^{-\Lambda}+\int\limits_{\Lambda}^{\Lambda_0}\right]
d\om \frac{\rho_0(\om)}{|\om|}.
\eeq
When $\Lambda\gg \Delta$, the integral in Eq. (\ref{rg}) averages over many peaks in the LDOS $\rho_0$ and
we obtain the usual result for the Kondo model with a constant density of states. 
It implies that in the limit $T\gg \Delta$, the result in Eq. (\ref{conduc1}) remains valid independently 
of $\rho_0(\omega)$
being tuned on or off a resonance.

On the other hand, 
when $\Lambda\ll \Delta$, the integral in Eq. (\ref{rg}) becomes strongly dependent on $\rho_0$. 
When $\rho_0$ is tuned off a resonance and for $\Delta\gg T\gg T_K^{NR}$,
the result in Eq. (\ref{conduc1}) remains approximately valid provided we replace 
$T_K^0$ by $T_K^{NR}$ defined in Eq. (\ref{def:tknr}) such that $G_{L0}(T)=G_{U} f(T/T_K^{NR})$.
When $\rho_0$ is tuned on a resonance that we assume for convenience to be at $\om=E_F=0$, 
only the variation of the LDOS in the vicinity of $E_F$ matters such that we further 
approximate the LDOS using Eq. (\ref{gamma0}).
The renormalization group equation in Eq. (\ref{rg}) may be  integrated in two steps, first 
between $\Lambda_0$ and $\Delta$ where the variations of the LDOS average out and then between $\Delta$ and $\Lambda$
where
\beq
J_{L0}(\Lambda)\approx J_{L0}(\Delta)+\frac{1}{2}J_{L0}(\Delta)J_{00}(\Delta)\rho_0(0)
\ln\left(1+\frac{\ga_n^2}{\Lambda^2}\right),
\eeq
where we assumed $\ga_n\ll \D$.
At $\Lambda\gg \Delta$, the Kondo couplings renormalize following the RG equations given in Eq. (\ref{RG0}). At the scale $\Delta$, the Kondo couplings have been weakly renormalized and one may use
the RG equations given in Eq. (\ref{RG0}) that imply
$J_{L0}(\Delta)\approx \frac{t_l t_0}{t_0^2} J_{00}(\Delta)$.
Using this approximation and the definition of the on-resonance Kondo temperature in Eq. (\ref{def:tkr}),
the conductance at $\D>T\gg \ga_n$ reads
\beq\label{conduc2}
G_{L0}\approx G_U \frac{3\pi^2}{16}\frac{1}{\ln^2(1+\frac{\ga_n^2}{(T_K^R)^2})}\frac{\ga_n}{T}
\left(1+\frac{\ln(1+\frac{\ga_n^2}{T^2})}{\ln(1+\frac{\ga_n^2}{(T_K^R)^2})}
\right)^2.
\eeq
The high temperature on-resonance conductance takes a more complicated form that
the one given in Eq. (\ref{conduc1}).
Notice nonetheless
that the multiplicative factor $(1+\cdots)$ appearing in Eq. (\ref{conduc2}) is $O(1)$
since $T\gg T_K^R=O(\ga)$.

\subsection{Scaling analysis}
From the low and high temperature analysis, we have seen that the conductance matrix elements 
cannot be in general simply written as a simple universal scaling function of $T/T_K^0$
both at high and low temperature. 
This is particularly 
striking when $\rho_0$ was tuned on a resonance characterized by the width $\gamma_n$
(see Eqs (\ref{glowt}) and (\ref{conduc2})).

In general, one would expect 
\beq\label{gscaling}
G_{\al\be}=G^U_{\al\be}~g(\frac{T}{T_K^0},\frac{\Delta}{T_K^0},\frac{\ga_n}{T}).
\eeq

In the high temperature regime, this scaling function takes a simple form and simply
reads $f(T/T_K)$ where $T_K$ either equates $T_K^0$ when $\Delta\ll T_K^0$ or can be simply 
expressed as a function of $T_K^0,\Delta,\gamma_n$ (see eqs (\ref{def:tkr}) and (\ref{def:tknr})).
Nevertheless, at intermediate or low temperature, 
the conductance in Eq. (\ref{gscaling}) takes a complicated scaling form 
which depends on $\ga_n/T$ and is in this sense non universal.
It depends on the geometric details of the sample, at least for bare Kondo temperature $T_K^0
$ of the order or smaller than the level spacing $\Delta$.
A similar conclusion has been reached by Kaul et al. \cite{kaul} by 
assuming a chaotic large quantum grain and explicitly calculating
the fluctuations and deviations from the universal behavior taking into account mesoscopic fluctuations.

\subsection{Finite grain Coulomb energy}
In this section, we discuss whether  a finite grain Coulomb energy  modifies
or not the results presented in this work. As we already mentioned in section 2, the Kondo
coupling $J_{00}$, is almost not affected by the grain Coulomb energy $E_G$ (since $E_G\ll U$)
and therefore the Kondo temperature remains almost unchanged.
As shown in \cite{oreg,borda04,florens},  a small  energy scale $E_G$ changes the 
renormalization group equation
in Eq. (\ref{RG0}). The off-diagonal couplings $J_{0L}(\Lambda),J_{0R}(\Lambda)$ tend to $0$ 
for $\Lambda\ll E_G$.
At energy $\Lambda\ll E_G$ 
the problem therefore reduces to an anisotropic $2-$channel Kondo problem. 
The strongly coupled channel is the grain $0$, the weakly coupled one is the even combination
of the conduction electron in the left/right leads. At very low energy, the fixed 
point of the anisotropic $2-$channel Kondo model is a Fermi liquid. It 
is characterized by the strongly coupled lead (here the grain)
screening the impurity whereas the weakly coupled one completely decouples from the impurity.
The dot density of states depicted in Fig. \ref{Fig:rhod} remains therefore almost unaffected.
The on and off-resonance Kondo temperatures $T_K^R$ and $T_K^{NR}$ given in Eqs (\ref{def:tkr}) and (\ref{def:tknr}) remain valid too.
The problem is to read the dot LDOS with the weakly coupled leads
since they decouple at $T=0$. Nevertheless, for a typical experiment done at low
temperature $T$, such a decoupling is not complete and the dot density of state should be still 
accessible using the weakly coupled leads but with a very small amplitude.

In this paper we analyze the situation in which a grain or a finite size wire is also
used as a third terminal. In some situations, like the theoretical one presented 
in [\onlinecite{oreg}], no terminal lead is attached to the grain
and the geometry is a genuine 2-terminal one. Taking into account both a finite level spacing and a finite grain Coulomb energy is quite involved (with several different regimes) and goes
beyond the scope of the present paper. A step into this direction was recently achieved in Ref.
[\onlinecite{kaul06}].

\section{Conclusions}

In this paper, we have studied a geometry in which a small quantum dot in
the Kondo regime
is strongly coupled to a large open quantum dot or open finite size wire
 and weakly coupled to other normal leads which are simply used as transport probes.
The artificial impurity is mainly screened is the large quantum dot. Such a geometry 
thus allows to probe the dot spectroscopic properties without perturbing it. 
We have shown combining several techniques how finite size effects show up in 
the dot density of states and in the the conductance
matrix. 
We hope the predictions presented here are robust enough to be 
checked experimentally.

\acknowledgments
This research was partly supported by the Institut de la Physique de la Matière Condensée (IPMC,Grenoble).


\begin{thebibliography}{}
\bibitem{note0} Note that for a $2D$ or $3D$ metallic grain, the level spacing $\Delta$ is not simply related to the Fermi velocity and the 
length scale of the grain. In this case, we shall compare directly $T_K^0$ to $\Delta$.
\bibitem{note1} This matrix has only
one non-zero eigenvalue, $Tr(\hat \lambda)\approx \lambda_{00}$ which flows to strong coupling. 
It follows that the dimensionless Kondo coupling
$\lambda_{\al\be}(\Lambda)=\frac{t_\al t_\be}{\sum_\al t_\al^2} Tr(\hat \lambda)(\Lambda)$ \cite{simon02}.
\end{thebibliography}
\end{document}